\documentclass[aps,prl,preprint,floatfix,superscriptaddress]{revtex4}

\usepackage{graphics}
\usepackage{epsfig}
\newcommand{\beq}{\begin{eqnarray}}
\newcommand{\eeq}{\end{eqnarray}}
\newcommand{\ket}[1]{|#1\rangle}
\newcommand{\bra}[1]{\langle#1|}

\begin{document}

\title
{Theory and {\it ab initio} calculation of radiative lifetime of excitons in
semiconducting carbon nanotubes} 
\author{Catalin D. Spataru} 
\affiliation{Department of Physics, University of California at Berkeley,
Berkeley, CA 94720, USA}
\affiliation{Materials Sciences Division, Lawrence Berkeley National
Laboratory, Berkeley, CA 94720, USA}
\author{Sohrab Ismail-Beigi}
\affiliation{Department of Applied Physics, Yale University, New
Haven, CT 06520, USA}
\author{Rodrigo B. Capaz}
\affiliation{Department of Physics, University of California at
Berkeley, Berkeley, CA 94720, USA}
\affiliation{Materials Sciences Division, Lawrence Berkeley National
Laboratory, Berkeley, CA 94720, USA}
\affiliation{Instituto de F\'\i sica, Universidade Federal do Rio de Janeiro, Caixa Postal 68528, Rio de Janeiro, RJ 21941-972, Brazil}
\author{Steven G. Louie}
\affiliation{Department of Physics, University of California at Berkeley,
Berkeley, CA 94720, USA}
\affiliation{Materials Sciences Division, Lawrence Berkeley National
Laboratory, Berkeley, CA 94720, USA}

\date{\today}

\begin{abstract}
We present theoretical analysis and first-principles calculation of
the radiative lifetime of excitons in semiconducting carbon nanotubes.  
An intrinsic lifetime of the order of $10$ ps is computed for the lowest 
optically active bright excitons.  The intrinsic lifetime is however a 
rapid increasing function of the exciton momentum.  Moreover, the 
electronic structure of the nanotubes dictates the existence of dark 
excitons  nearby in energy to each bright exciton.  Both effects 
strongly influence measured lifetime.  Assuming a thermal occupation 
of bright and dark exciton bands, we find an effective lifetime of 
the order of $10$ ns at room temperature, in good accord with recent 
experiments.
\end{abstract}
\maketitle
Single-walled carbon nanotubes (SWCNT) possess highly unusual 
mechanical, thermal,
electronic and optical properties, making them objects of great
interest for basic scientific studies as well as potential
applications \cite{Industrial}.
Their optical properties have been studied intensively
in the past, but most of the optical measurements were
performed on aggregated samples containing bundles of SWCNT
\cite{Kataura,Ichida}, resulting
in measured spectra broadened or even washed out by environmental
interaction effects and preventing a detailed spectral analysis. 
Recently,  fabrication of well-separated SWCNT
\cite{OConnell,Lefebvre} 
allowed for much better resolved optical absorption spectra,
and more importantly fluorescence was detected from individual
semiconducting SWCNT. Combining resonant fluorescence and
Raman spectroscopy, it became possible to assign optical
transitions to specific $(n,m)$ semiconducting SWCNT
\cite{Bachilo,MIT}.

While first-principles calculations \cite{Spataru1,Spataru2,Chang} have
explicitly predicted huge exciton binding energies (of the order of 1
eV) for semiconducting tubes with prominent features in their 
optical spectra excitonic in nature and there were also several model
studies \cite{Ando,Perebeinos,Mazumdar}, most experimental data
have been interpreted by assuming one-electron 
inter-band transitions. Thus, any information on photoexcited
states of SWCNT is very valuable. In particular, knowledge of the
radiative lifetime would be helpful in understanding the nature of 
the excitations and essential for potential applications in photonics
and optoelectronics.

Recent time-resolved flourescence experiments reported a 
radiative lifetime of excited nanotubes ranging from $10$ ns
\cite{Hagen} 
to $100$ ns \cite{Wang}. The variation in the measured radiative
lifetime is probably due
to uncertainty in the quantum efficiency of fluorescence of the semiconducting SWCNT that contribute to the fluorescence signal.

In this work, we carry out {\it ab initio} calculations of the
radiative lifetime of the lowest electronic excitations in several 
semiconducting SWCNTs. The electronic excited states are found by
solving the Bethe-Salpeter equation, including 
band-structure, self-energy, and electron-hole interaction effects
\cite{Hybertsen,Rohlfing,Spataru1,Spataru2}.   Solution of the Bethe-Salpeter equation provides the excitation energies and quantum amplitudes (or wave functions) for excitons.  Specifically, for an excited state with center of mass momentum $\hbar \vec Q$ labeled by index $S$, we have the excitation energy 
$\hbar \Omega_S(\vec Q)$ and the amplitude
\beq
\chi(\vec r_e,\vec r_h) = \bra{G}\Psi(\vec r_h)^\dag\Psi(\vec r_e)\ket{S(\vec{Q})}\,.
\eeq
Here, $\Psi(\vec r)^\dag$ and $\Psi(\vec r)$ are creation and annihilation field operators that create or destroy an electron at position $\vec r$.  The square amplitude $|\chi(\vec r_e,\vec r_h)|^2$ is the probability density of finding the electron at $\vec r_e$ and hole at $\vec r_h$.  Within the Tamm-Damcoff
approximation \cite{Fetter}, the amplitude takes the form of a coherent superposition of electron-hole pairs,
\beq
\chi(\vec r_e,\vec r_h) = \sum_{vc\vec{k}}
A^{S(Q)}_{vc\vec{k}}\, 
\psi_{c\vec{k}+\vec{Q}}(\vec r_e)\, \psi_{v\vec{k}}(\vec r_h)^*\,.
\eeq
Here, $\psi_{n\vec k}(\vec r)$ are Bloch states with wave vector $\vec
k$, and $c$ and $v$ label conduction (electron) and valence (hole)
bands.  The coefficients $A^{S(Q)}_{vc\vec{k}}$ which describe the
weight of electron-hole pairs in the coherent superposition are
obtained from the Bethe-Salpeter equation along with the excitation
energy $\hbar \Omega_S(\vec Q)$.

\begin{figure}
\resizebox{12.0cm}{!}{\includegraphics{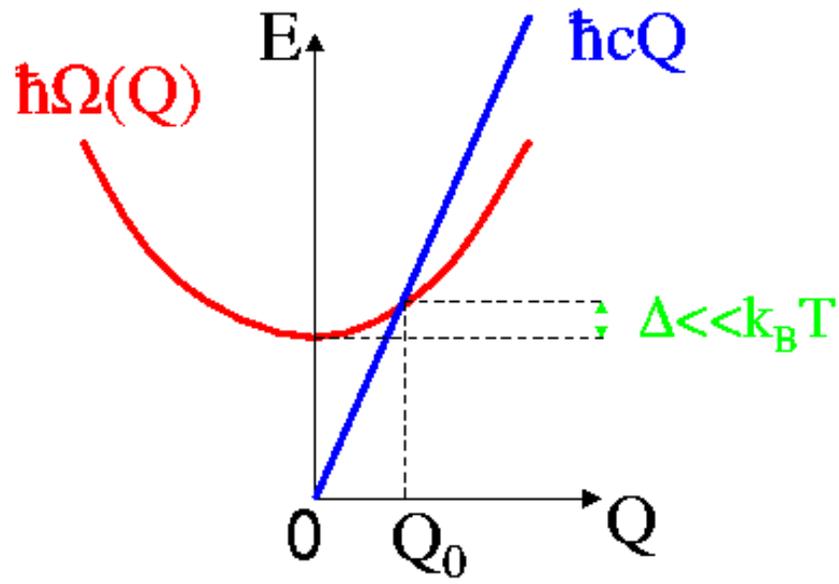}}
\caption{Schematic view of an exciton band as a function of center of
mass momentum and the energy of a photon
with momentum parallel to the tube axis; $Q_0$ is the maximum momentum
of an exciton which can still decay through emission of a photon.}
\label{bright}
\end{figure}

The radiative lifetime is calculated using Fermi's golden rule.
We start with the usual Hamiltonian describing the
interaction between electrons and photons,
\beq
H^{int} & = & -\frac{e}{mc} \int d\vec{r}~ \Psi(\vec{r})^\dag
\vec{A}(\vec{r}) \cdot  \vec{p} ~\Psi(\vec{r})\,,
\eeq
where $\vec{p}=-i\hbar\vec\nabla$ and $\vec{A}(\vec{r})$ is the
vector potential for the electromagnetic field in the Coulomb gauge.  We decompose $\vec{A}(\vec{r})$ into plane wave modes \cite {Loudon}:
\beq
\vec{A}(\vec{r}) = \sum_{\vec q \lambda} \sqrt{\frac{2\pi\hbar c}{V q}}
~\vec{\epsilon}_{\vec q \lambda}\left[a^{\dag}_{\vec q \lambda}
e^{-i\vec q \cdot \vec r} + h.c. \right]\,,
\eeq
with $a^{\dag}_{\vec q \lambda}$ ($a_{\vec q \lambda}$) being the  
creation (annihilation) operators of photons with polarization vector
$\vec\epsilon_{\vec{q}\lambda}$. The radiative decay rate for a system
in an initial
state with one exciton and no photon $\ket{S(\vec{Q}),0}$ to the
electronic ground state is given by:
\beq
\gamma(\vec Q) = \frac{2\pi}{\hbar} \sum_{\vec q \lambda} 
\left| \langle G, 1_{\vec q \lambda} |H^{int}| S(\vec Q), 0 \rangle
\right|^2 \delta (\hbar \Omega(\vec Q) - 
\hbar cq)\,.
\eeq
with
\beq
\bra{G, 1_{\vec q \lambda}}H^{int}\ket{S(\vec Q), 0} 
 =  -\frac{e}{mc}\sqrt{\frac{2\pi\hbar c}{Vq}}\, \vec\epsilon_{\vec{q}\lambda}\cdot
\sum_{vc\vec{k}} A^{S(Q)}_{vc\vec{k}}\,
\bra{v\vec{k}}e^{-i\vec q\cdot \vec r}\vec{p}\ket{c\vec{k}+\vec{Q}}\,.
\eeq
In a quasi-1D system like a SWCNT, momentum conservation
requires the photon momentum along the tube axis $\hat{z}$
to equal the exciton momentum $\vec{Q}=Q\hat{z}$. And for optical photon
energies, we may use the dipole approximation:
\beq
\bra{v\vec k}e^{-i\vec q\cdot \vec r}\vec{p}\ket{c\vec k +\vec Q}
\approx
\bra{v\vec k}\vec{p}\ket{c\vec k} 
\times \delta_{q_z,Q}\,.
\eeq
Moreover, SWCNTs have negligible optical response for electric fields 
polarized perpendicular to the tube axis, due to the 'depolarization
effect' \cite{Ajiki}. Thus,
\beq
\vec{\epsilon}_{\vec q \lambda}
\cdot \langle G| \vec{p}| S(0)\rangle \approx \vec{\epsilon}_{\vec q \lambda}
\cdot \hat z ~\langle G|p_z| S(0)\rangle,
\eeq
where we have made use of the  shorthand definition:
\beq
\bra{G}\vec{p}\ket{S(0)} = \sum_{vc\vec k}A^{S(0)}_{vc\vec k}
\bra{v\vec k}\vec{p}\ket{c\vec k}\,.
\eeq
Summing over two arbitrary orthogonal polarizations of the electric
field, and increasing the volume $V=L_xL_yL_z$ to $\infty$, 
the decay rate becomes:
\beq
\gamma(Q) = 
\lim_{L_z\rightarrow\infty}  \frac {\left | \langle G|p_z| S(0)\rangle
\right |^2}{L_z} \frac{e^2}{m^2c^2\hbar}
\int_{-\infty}^{\infty} d q_{x} \int_{-\infty}^{\infty} d q_{y}
\nonumber \\ \times
\left[\frac{q_x^2+q_y^2}{(q_x^2+q_y^2+Q^2)^{3/2}}
\right] 
{\delta{\left( \frac{\Omega(Q)}{c} -
\sqrt{q_x^2+q_y^2+Q^2}\right)}}
\eeq
In our calculations, finite sampling of N k-points for the above
equations is equivalent to have a supercell along the tube axis with 
$L_z=Na$, where $a$ is the physical unit cell size along the tube axis.
Using the relation of momentum and velocity,
$p_z=m v_z=m \frac{i}{\hbar}\left[H,z\right]$,
\beq
\lim_{L_z\rightarrow\infty}  \frac {\left | \langle G|p_z| S(0)\rangle
\right |^2}{L_z} = 
m^2\Omega(0)^2
\lim_{N\rightarrow\infty} \frac {\left |
\langle G|z| S(0)\rangle \right |^2}{Na} 
\equiv  {m^2\Omega(0)^2}  \frac{\mu_a^2}{a}\,,
\eeq
where $\frac{\mu_a^2}{a}$ is the squared exciton transition dipole
matrix element per unit tube length. Performing the integral in eq. (10),
the decay rate can finally be written as:
\beq
\gamma(Q) = \left \{
\begin{array}{cc}
\frac{2\pi e^2 \Omega(0)^2}{\hbar c^2} \frac{\mu_a^2}{a} 
\frac{{\Omega^2(Q)-c^2Q^2}}{\Omega^2(Q)} & $~~~~~~if~$ ~|Q| \leq {Q_0} \\
0 &  $~~~~~~if~$ ~|Q| \geq {Q_0}
\end{array}\right.\,,
\eeq
where $Q_0$ defined by the condition,
$\hbar \Omega(Q_0) = \hbar {\textstyle c} Q_0\,$, is the maximum
momentum a decaying exciton can have. The decay of excitons with 
momentum $Q > Q_0$ is forbidden by momentum conservation: such a decay 
would have the photon momentum along $\hat z$ be smaller than the 
exciton momentum. 
From Fig. \ref{bright} we see that for optical exciton energies where
the photon momentum is small compared to the scale of Brillouin zone,
$Q_0 \approx
\frac{\Omega(0)}{c}\,$.
Thus the intrinsic radiative 
lifetime $\tau$ of an exciton is
shortest for an exciton with $Q=0$ and
increases as a function of momentum until it becomes infinite for 
$Q= Q_0$ \cite{Chen}. 

The above expression for the radiative decay rate of a 
zero-momentum exciton in a SWCNT,
$\gamma(0)=\frac{2\pi e^2 \Omega(0)^2}{\hbar c^2} \frac{\mu_a^2}{a}
$,
may be compared to
that of a molecule of size $l$,
$\gamma=\frac{4 e^2 \Omega^3}{3 \hbar c^3} {\mu^2}\,$.
Assuming similar excitation energies $\hbar \Omega(0) \sim \hbar\Omega$ and
squared dipole moments per unit length $\frac{\mu_a^2}{a} \sim
\frac{\mu^2}{l}$
for excitons in SWCNT and  molecule, comparison of the two expressions indicates 
that $\gamma(0)$ is enhanced with respect to $\gamma$ by a factor 
$\frac{\lambda}{l}$. 
Thus we expect the intrinsic radiative lifetime of zero-momentum excitons in a SWCNT to be about $3$ orders of magnitude shorter than the
radiative lifetime of  excitons in small molecules, which is typically on the
order of a few ns.
Table \ref{table1} shows our {\it ab initio} results for $\frac{\mu_a^2}{a}$ 
and the intrinsic radiative 
lifetime of the lowest bright exciton $\tau(0)$ 
for the four semiconducting
SWCNT we considered: (7,0), (8,0), (10,0) and (11,0). While 
$\frac{\mu_a^2}{a}$ is indeed of the order of a few $a.u.$, $\tau(0)$ is
much shorter than the typical radiative
lifetime in small molecules.
We find $\tau(0)$ between 8 ps and 19 ps for zigzag tubes with diameters between 5.5 \AA ~and 8.7 \AA. We see well-defined diameter and 
chirality trends in the results of Table \ref{table1}: $\tau(0)$ increases with diameter 
and shows a $\nu$-family oscillation, where $\nu = (n-m)$ mod 3. Such oscillations 
usually reveal chirality dependences in the form of $(-1)^\nu
\cos(3\theta)$ \cite{Capaz}, where $\theta$ is the chiral angle. 

Because the intrinsic radiative lifetime
of bright excitons ranges from $\sim 10$ ps to $\infty$ depending on
their momentum $Q$, we use an exciton distribution
function to estimate an effective radiative lifetime
$\tau^b_{_{eff}}$. Experimentally, $\tau_{_{eff}}$ has been deduced from
time-resolved fluorescence data taken over a time range of several
tens of ps \cite{Hagen,Wang}. 
We assume that much more rapid relaxation processes exist in SWCNT and
make the excitons assume a thermal 
(Boltzman) distribution. We employ an effective mass
approximation for the exciton energy dispersion:
$\hbar \Omega(Q) = \hbar \Omega(0) + \frac{\hbar^2 Q^2}{2 M}$,
with $M  = m^*_e +m^*_h$ being simply the sum of the effective 
electron and hole masses of the 
conduction and valence bands which comprise the exciton.
The effective masses $ m^*_e$ and $m^*_h$ are taken from
DFT (density functional theory) band-structure calculations. 
We are thus negleting two
different effects which tend to cancel each other: the quasiparticle
self-energy effects which would give electron and hole  
 masses smaller than the values predicted by DFT, and the
electron-hole interaction effects which would give an exciton mass
larger than the sum of the electron and hole masses \cite{Mattis}.
With these assumptions, the effective decay rate of bright excitons $\gamma^b_{_{eff}}$ can be estimated by performing
a simple statistical average \cite{Citrin}:
\beq
\gamma^b_{_{eff}} 
= \gamma(0)\cdot
\frac{\int_0^\Delta dE {\frac{1}{\sqrt{E}}
e^{-\frac{E}{k_BT}}}~
 \frac{\left[\hbar\Omega(0)+E\right]^2-2M c^2 E} {[\hbar\Omega(0)+E]^2}}
{\int_0^\infty dE {\frac{1}{\sqrt{E}} e^{-\frac{E}{k_BT}}}}
\eeq
where $E=\frac{\hbar^2 Q^2}{2 M}$ is the energy of an exciton
measured from the bottom of the exciton band and $\Delta$ is the maximum
energy $E$ that a radiatively decaying exciton can have (see
Fig. \ref{bright}). Given the smallness of the photon momentum, one has
$\Delta = \frac{\hbar^2 Q_0^2}{2 M} \approx \frac{\hbar^2\Omega^2(0)}{2
Mc^2}  \ll k_BT\,$. Evaluation of eq. (13) at finite temperature gives:
\beq
\gamma^b_{_{eff}} 
\approx \frac{4}{3} \frac{1}{\sqrt{\pi}}
\sqrt{\frac{\Delta}{k_BT}}\times\gamma(0)\,.
\eeq
Thus, the effective radiative lifetime  $\tau^b_{_{eff}}$ gets
enhanced with respect to the intrinsic radiative lifetime of a bright
exciton $\tau(0)$ by a factor of approximately
$\sqrt\frac{k_BT}{\Delta}$. At room temperature this factor is
$\approx 100$, reflecting the fact that $\Delta \ll k_BT$: only a small fraction of thermalized excitons are in the allowed radiative region.
In Table \ref{table2} we present our {\it ab initio} results for 
the effective mass $M$ and 
the effective radiative lifetime $\tau^b_{_{eff}}$ of the lowest
bright exciton at $T=300$ K for four semiconducting SWCNT considering
only the intra-exciton-band distribution effect. At this level of
analysis we see that  $\tau^b_{_{eff}}$ is of the order of about $1$ ns at 
room temparature for the SWCNT we studied, with larger values for the
larger diameters.

\begin{figure}
\resizebox{12.0cm}{!}{\includegraphics{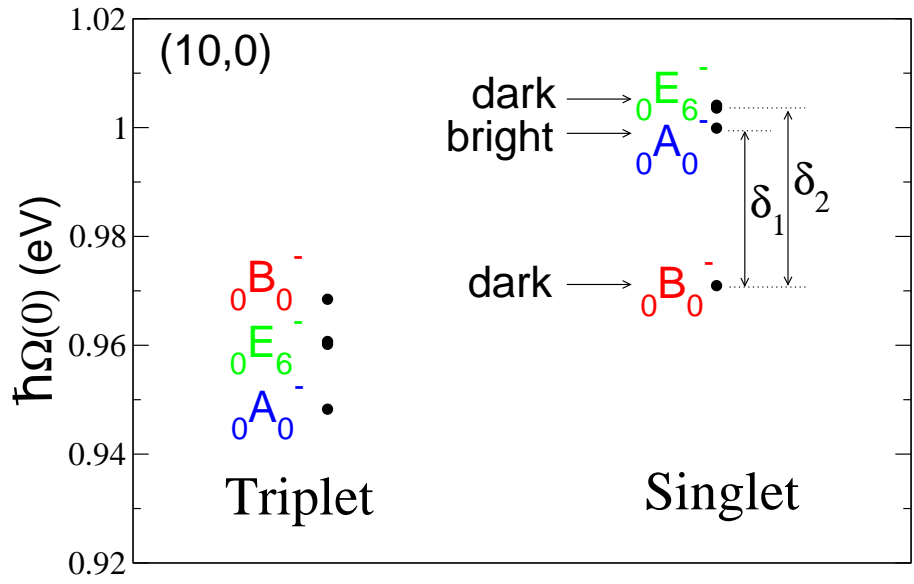}}
\caption{Excitation energies of the lowest spin-triplet and
spin-singlet excitons with zero momentum for the (10,0) SWCNT. }
\label{splitting}
\end{figure}

Until now we considered only bright excitons when evaluating
the effective radiative lifetime. However, a more accurate description of
$\tau_{_{eff}}$ should take into
account the complete electronic structure of the lowest energy complex
of spin-singlet
excitonic states, which includes dark states as well 
\cite{Spataru1,Spataru2,Mazumdar}. Figure \ref{splitting} shows our calculated 
electronic structure
of the lowest excitonic states in the (10,0) tube (a similar picture
holds for (11,0)). The states are
labeled according to the irreducible representation of the
symmetry group to which they belong \cite{Vuk}. 
We see that the lowest singlet exciton ($_0B_0^-$) is actually a dark one, 
followed by a bright ($_0A_0^-$) and then two other ($_0E_{6}^-$) degenerate 
dark
excitons. The key observation is that both the optical transition matrix 
element and the exciton exchange energy 
depend on the exciton amplitude $\chi(\vec r,\vec r)$ evaluated for
electron and hole at the same position. Excitons of $_0B_0^-$ are 
odd under $\sigma_v$ (vertical planes) reflections and therefore have 
$\chi(\vec r, \vec r)=0$ on these planes \cite{futurework}.  These nodes and
concomitant oscillations in sign of $\chi$ lead to a vanishingly small exchange
energy.  Figure \ref{splitting} shows that the exchange interaction shifts up 
all other singlet excitons states with respect to their triplet
counterparts and the $_0B_0^-$ then becomes the lowest singlet
state. A similar analysis can be made for chiral tubes. In fact, our {\it ab 
initio} results, in combination with a symmetry analysis based on
group theory, show in general that the lowest exciton state is dark 
for any zigzag and chiral semiconducting SWCNT \cite{futurework}, as
long as their diameter is not very small \cite{70-80}.
{\it Therefore, optical darkness and small exchange energies have the same 
physical origin, a 
fundamental result for the understanding of the low efficiency of light 
emission in SWCNT.}

\begin{figure}
\resizebox{12.0cm}{!}{\includegraphics{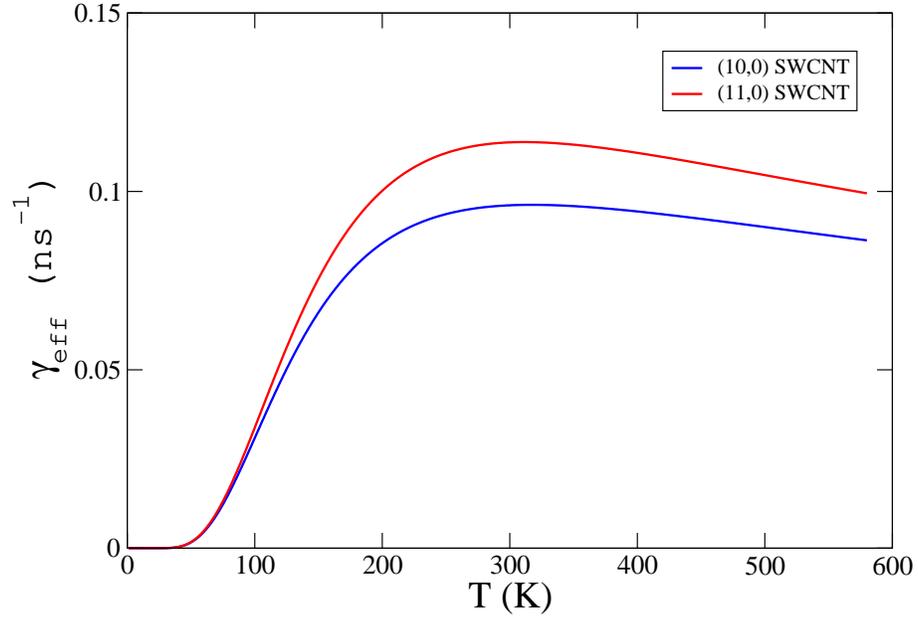}}
\caption{Temperature dependence of the inverse effective radiative
lifetime $\gamma_{_{eff}}$ for the excited nanotubes (10,0) and (11,0), taking
into account thermal occupation of dark and bright exciton states.}
\label{temp_dep}
\end{figure}

Now, let's assume that there is perfect thermalization between dark and
bright excitonic bands, given the fact that their energy separation is
 $\sim k_BT$ at room temperature. Then, it is easy to see that the
statistical average will result in an effective radiative decay rate
$\gamma_{_{eff}}$:
\beq
\gamma_{_{eff}}=
\frac{e^{-\delta_1/k_BT}}
{1+e^{-\delta_1/k_BT}+2e^{-\delta2/k_BT}}
\times \gamma^b_{_{eff}}
\eeq
where the parameters $\delta_1$ and $\delta_2$ are defined in
Fig. \ref{splitting}. The dark excitons further enhance the effective radiative
lifetime of excited SWCNT. Our {\it ab initio} results for $\delta_1$,
$\delta_2$ and $\tau_{_eff}$ in the (10,0) and (11,0) tubes, shown in 
Table \ref{table2}, indicate that this enhancement factor is $\sim 5$
at room temperature,
bringing the effective radiative lifetime to an order of magnitude of $\sim
10$ ns, in  good agreement with experimental results. The overall
temperature dependence of the effective radiative lifetime within this
framework is illustrated in Fig. \ref{temp_dep}.

In summary, we have presented a theory and calculation, based on {\it
ab initio} results, of the intrinsic radiative lifetime of 
bright excitons
in several semiconducting single walled carbon nanotubes. Taking account of
dark excitons and assuming
fast thermalization processes among both bright and dark excitons,
our results provide an 
explanation for the effective radiative lifetime of excited
nanotubes measured at room temperature.

This work was supported by the NSF
under Grant No. DMR04-39768, and the U.S. DOE under Contract
No. DE-AC03-76SF00098. Computational resources have been provided by
NERSC and NPACI. R.B.C. acknowledges financial support from the
Guggenheim Foundation and Brazilian funding agencies CNPq, CAPES, FAPERJ, Instituto de Nanoci{\^e}ncias, FUJB-UFRJ and
PRONEX-MCT.

\clearpage
\begin{table}
\caption{The dipole matrix element per unit length, excitation energy and
intrinsic radiative lifetime of 
the lowest bright exciton ($Q=0$) for several zig-zag SWCNT.}
\label{table1}
\begin{tabular}{cccc}
\hline\noalign{\smallskip}
SWCNT & $\frac{\mu_a^2}{a}(a.u.)$ &$\hbar \Omega$(0)$(eV)$ & $\tau$(0)$(ps)$  \\
\noalign{\smallskip}\hline\noalign{\smallskip}
(7,0)  & 2.8  & 1.20 &  12.8 \\
(8,0)  & 2.8   & 1.55 &  8.1  \\
(10,0)  & 2.8  & 1.00  & 19.1 \\
(11,0)  & 2.5 & 1.21  & 14.3 \\
\noalign{\smallskip}\hline
\end{tabular}
\end{table}

\begin{table}
\caption{The effective mass, effective lifetime at room 
temperature neglecting dark excitons $\tau^b_{_{eff}}$, 
dark-bright exciton splitting parameters $\delta_1$ and $\delta_2$,
and the effective radiative lifetime at room
temperature including dark excitons $\tau_{_{eff}}$. (See text.)}
\label{table2}
\begin{tabular}{cccccc}
\hline\noalign{\smallskip}
SWCNT & M$(m_e)$ & $\tau^b_{_{eff}}(ns)$& $\delta_1$(meV) & $\delta_2$(meV) & $\tau_{_{eff}}(ns)$  \\
\noalign{\smallskip}\hline\noalign{\smallskip}
(7,0)  & 0.21  & 1.1  & - & - & -  \\
(8,0)  & 0.71  &  1.0  & - & - & - \\
(10,0)  & 0.19  &  1.8  & 29 & 33  &  10.4 \\
(11,0)  & 0.44 & 1.7  & 29 & 44 & 8.8 \\
\noalign{\smallskip}\hline
\end{tabular}
\end{table}

\clearpage

\clearpage

\clearpage

\end{document}